\documentclass[aps,pra,superscriptaddress,12pt,tightenlines,nofootinbib]{revtex4}
\usepackage{amsmath,amsthm,graphicx,amssymb,verbatim,listings}
\usepackage{wasysym}
\usepackage[T1]{fontenc}     
\usepackage{lmodern}         
\usepackage[ pdftex, plainpages = false, pdfpagelabels,
                 pdfpagelayout = useoutlines,
                 bookmarks,
                 bookmarksopen = true,
                 bookmarksnumbered = true,
                 breaklinks = true,
                 linktocpage=all,
                 pagebackref=false,
                 colorlinks = true,
                 linkcolor = BrickRed,
                 urlcolor  = blue,
                 citecolor = BrickRed,
                 anchorcolor = green,
                 hyperindex = true,
                 hyperfigures
                 ]{hyperref}
\usepackage[usenames, dvipsnames]{xcolor}
\usepackage{xifthen} 

\newcommand{\tr}{\hbox{tr}}

\newcommand{\ket}[1]{{\ensuremath{\left| #1 \right\rangle}}}

\newcommand{\arxiv}[2][]{\ifthenelse{\isempty{#1}}{\href{http://arxiv.org/abs/#2}{{\tt arXiv:\allowbreak{}#2}}} {\href{http://arxiv.org/abs/#2}{{\tt arXiv:\allowbreak{}#2 [#1]}}}}

\newcommand{\booktitle}{\textsl}
\newcommand{\hrefdoi}[2]{\href{https://dx.doi.org/#1}{#2}}

\begin{document}

\title{Ideas Abandoned en Route to QBism}

\author{Blake C.\ Stacey}
\affiliation{\href{http://www.physics.umb.edu/Research/QBism/}{Physics
    Department}, University of Massachusetts Boston}

\date{\today}
\begin{abstract}
The interpretation of quantum mechanics known as QBism developed out
of efforts to understand the probabilities arising in quantum physics
as Bayesian in character. But this development was neither easy nor
without casualties. Many ideas voiced, and even committed to print,
during earlier stages of Quantum Bayesianism turn out to be quite
fallacious when seen from the vantage point of QBism.
\end{abstract}
\maketitle

If the profession of science history needed a motto, a good candidate
would be, ``I think you'll find it's a bit more complicated than
that.''  This essay explores a particular application of that
catechism, in the generally inconclusive and dubiously reputable area
known as quantum foundations.

QBism is a research program that can briefly be defined as
\begin{quotation}
\noindent an interpretation of quantum mechanics in which the ideas of
\emph{agent} and \emph{experience} are fundamental. A ``quantum
measurement'' is an act that an agent performs on the external
world. A ``quantum state'' is an agent's encoding of her own personal
expectations for what she might experience as a consequence of her
actions. Moreover, each measurement outcome is a personal event, an
experience specific to the agent who incites it. Subjective judgments
thus comprise much of the quantum machinery, but the formalism of the
theory establishes the standard to which agents should strive to hold
their expectations, and that standard for the relations among beliefs
is as objective as any other physical theory~\cite{DeBrota:2018}.
\end{quotation}
The first use of the term \emph{QBism} itself in the literature was by
Fuchs and Schack in June 2009~\cite{Fuchs:2009}. Prior to this, they
had employed it in talks and
correspondence~\cite[p.\ 1707]{Fuchs:2014}, illustrating the
lexicographer's principle that words predate their preservation in
books. Introducing a collection of correspondence, Fuchs wrote that
three characteristics of the QBist research program distinguish it
from existing interpretations~\cite[p.\ ix]{Fuchs:2014}.
\begin{quotation}
  \noindent First is its crucial reliance on the mathematical tools of
  quantum information theory to reshape the look and feel of quantum
  theory's formal structure. Second is its stance that two levels of
  radical ``personalism'' are required to break the interpretational
  conundrums plaguing the theory. Third is its recognition that with
  the solution of the theory's conundrums, quantum theory does not
  reach an end, but is the start of a great journey.
\end{quotation}

QBism grew out of Quantum Bayesianism, a loosely-defined school of
thought typified by a paper by Caves, Fuchs and Schack, ``Quantum
probabilities as Bayesian probabilities'' (\cite{CFS:2002},
hereinafter CFS 2002).  The most prominent thread within Quantum
Bayesianism may be the one that, in the hands of Fuchs and Schack,
later developed into QBism, but the term is applicable much more
broadly. It encompasses some writings of Bub and
Pitowsky~\cite{Pitowsky:2002, Bub:2009, Bub:2019}, for example, and
could easily include earlier suggestions of
Youssef~\cite{Youssef:1994} and Baez~\cite{Baez:2003}, work by Leifer
and Spekkens~\cite{Leifer:2013}, the ``entropic dynamics'' of
Caticha~\cite{Caticha:2007} and so forth.  These views overlap to the
extent that they all advocate interpreting the probabilities in
quantum physics according to some variety of Bayesianism.  They
overlap, but they do not coincide. What else should one expect, given
that even before quantum physics was brought into the game, the jest
that there were ``46,656 varieties of Bayesianism''~\cite{Good:1983}
was only a mild exaggeration?

Perusing the quantum-foundational literature and partaking in
conversations with quantum foundationers, I have been surprised by the
inclination to cite CFS 2002 as defining QBism. Sometimes, it is
invoked by itself as the canonical QBist document, and on other
occasions, it is mixed together indiscriminately with genuinely QBist
sources. Without commenting in depth on the merits of these varied
works, I find this situation puzzling. All three authors disavow the
perspective of this paper~\cite{CFS:2007, Fuchs:2009}, whether they
have continued along the path of developing QBism (Fuchs and Schack)
or not (Caves).

The philosophy of science, we fondly imagine, should be the discipline
that exposes the distinctions which physicists coarsely gloss
over. Perhaps QBism has generated too many expositions to pay
attention to just one~\cite{Fuchs:2013c, Fuchs:2017, Fuchs:2017b,
  Fuchs:2019}, but reading is in the scholar's job description, and
fortunately, not all QBist writings are as long as some of them
are. Indeed, some authors have gotten the coordinates right for
genuine QBist expositions~\cite{Cabello:2017, Frauchiger:2018,
  Koberinski:2018, Schaffer:2019}.

Space constraints and the mostly ahistorical writing style of physics
journals have prevented earlier QBist articles from delineating which
Quantum Bayesian papers still have valuable portions, which are nearly
obsolete, which are by authors who may sympathize with QBism without
fully subscribing to it, and so forth.  (These constraints have also
inhibited enumerating those articles which fail to distinguish CFS
2002 from genuinely QBist sources, which ones seem to ignore all more
recent developments and take CFS 2002 as the latest word in
``informational interpretations'', etc.)  And among the more leisurely
portrayals of QBism, the book by von Baeyer~\cite{VonBaeyer:2016} was
pitched to the general pop-science audience, making it ill-suited to
address the ``inside baseball'' matters like different schools of
Bayesianism.  Apart from a brief note in the context of a trendy but
confined discussion~\cite{Stacey:2019}, I myself have not drawn a hard
line between QBism and the more amorphous Quantum Bayesianism that
came before it, thinking in my innocence that the progression of
thought was dramatic enough that it did not need pointing
out. Correcting this deficit --- and atoning, in part, for my blithe
na{\"\i}vet\'e --- turns out to be an educational exercise. This is
the second time I have marshalled historical evidence to show that a
well-cited work was not in fact QBist, despite third-party claims to
that effect~\cite{Stacey:2016}. From the viewpoint of QBism's
$h$-index, this must appear a quixotic or even self-destructive
effort, but integrity is never easy.

\section{Locating QBism}

Statements about probability have been given many different
interpretations over the years.  One sect would read an equation like
``$p = 0.7$'' as a claim about relative frequency in a large ensemble,
while another would like to take it as concerning the extent to which
a proposition follows logically from evidence.  The Bayesian tradition
regards probabilities as quantities asserted by gamblers, and it is
within this tradition that QBism situates itself.  Each of these
intellectual genera contains many species.  Within Bayesianism, one
might mandate that in principle, all probabilities should reduce to 0
or 1 --- maximal and complete information must resolve all
uncertainty.  This is the spirit we find, for example, in Jaynes or
Garrett~\cite{Garrett:1993}.  Another aspiration has it that within
each physical situation, there dwells something like a chance density
or a ratio of up to down probabilitons, so that any gambler aware of
the value of that ``objective chance'' must set her odds to the exact
figure it implies.  This is difficult, perhaps impossible, to make
logically self-consistent or to integrate with known quantities in
physics; in day-to-day work, the postulation of such properties seems
extraneous to scientific practice~\cite{Fuchs:2019}.  QBism instead
follows the lead of \emph{personalist Bayesianism,} a view
historically associated with Ramsey and de Finetti~\cite{Ramsey:1926,
  Jeffrey:1989}.  It eschews the probabilitons and finds objectivity
at a different level.  For the QBist, no intrinsic attribute of a
physical system can itself compel the outcome of a quantum measurement
upon that system, nor even the probabilities that an agent should
ascribe to the potential outcomes of that measurement before she
performs it.

The research program of QBism does not content itself with providing a
story for the familiar mathematical formalism of quantum theory. Nor
is it satisfied with detailing how QBism differs from previous
attempts to interpret the quantum --- a harmless pastime for those who
treasure the peace of library basements.  Rather, the goal is to
understand why that formalism is useful:\ Why quantum theory, as
opposed to any alternative we might envision?  QBism finds the exhortation
to ``shut up and calculate!''\ unstable against perturbations by
curiosity:\ ``Were the world a different way, would we not, after we
shut up, calculate in a different fashion?''~\cite{Stacey:2019e}.

Three Greek-derived words are helpful in discussing interpretations of
quantum mechanics. \emph{Ontic} refers to entities and quantities that
exist, in their own right, in blunt reality --- in a Newtonian
worldview, the mass of a rock is ontic. \emph{Epistemic} quantities
have the character of knowledge, while \emph{doxastic}, from the Greek
for ``belief'', captures the personalist Bayesian view of
probabilities, and thus the QBist interpretation of quantum
states. Writing a wavefunction $\ket{\psi}$ is staking out a doxastic
claim, though the fact that it has proven useful to use vectors in
complex Hilbert spaces to express our doxastic statements has an
ontological lesson subtly coded within it.

QBism is largely orthogonal to matters of ``Bayesian inference'' as
understood in statistics or big-data science~\cite[\S 9]{DeBrota:2018}.
Attempting to grasp what QBism is about by extrapolating a Google
University education in those subjects has led more than a few poor
souls into confusion, whether they recognize it or not.

The remainder of this article will be devoted to identifying the
differences between QBism and what we might call ``proto-QBism'', the
views articulated by Fuchs, Schack and coauthors in the 1990s and
early 2000s. We will begin with the CFS 2002 paper mentioned above,
which in my informal experience has been most commonly confused with
QBism proper, and which provides a rather nice contrast with it.  We
will follow that with a close study of a follow-up article that Caves,
Fuchs and Schack wrote a few years later, which as we will see still
is not QBism. We will then backtrack and examine older writings of
Fuchs himself that are less widely invoked, and whose distance from
QBism is in certain aspects almost shocking. Our final exhibit will be
a position statement that Fuchs and Asher Peres wrote for
\booktitle{Physics Today}. My hope is that revealing this history may
help explicate why QBism developed as it did, and that it may aid
those displeased with QBism to be unhappy with QBism itself instead of
a confabulation.

\section{Caves--Fuchs--Schack (2002)}
CFS 2002 makes a case that quantum probabilites should be regarded as
Bayesian probabilities, but it doesn't do much more than that --- at
least, not very well, and from a QBist perspective, not convincingly.

The ceaseless and uncritical invocation of ``maximal information'' in
CFS 2002 is legitimately grating to a QBist ear. At best, one can
wincingly try to find a reading where it is tautological, treating the
statements \emph{Alice has maximal information about a system S} and
\emph{Alice ascribes a pure quantum state to system S} as wholly
synonymous. But one would be cruelly paid back for such
generosity. The years have taught us that it is just not possible to
shake the connotations of ``maximal information'' --- connotations of
pre-existing properties, of the ``ontic states'' that a more timid
view would want to underlie quantum theory. As Fuchs and Schack would
write much later~\cite{Fuchs:2009},
\begin{quotation}
  \noindent The trouble with the phrase ``maximal information is not
  complete'' and the imagery it entails is that, try as one might to
  portray it otherwise (by adding ``cannot be completed,'' say), it
  hints of hidden variables. What else could the ``not complete''
  refer to?
\end{quotation}
Therefore, we must read CFS 2002 through the prism of this later
repudiation of it. Section IV begins,
\begin{quotation}
  \noindent Our concern now is to show that if a scientist has maximal
  information about a quantum system, Dutch-book consistency forces
  him to assign a unique pure state. Maximal information in the
  classical case means knowing the outcome of all questions with
  certainty.  Gleason's theorem forbids such all-encompassing
  certainty in quantum theory. Maximal information in quantum theory
  instead corresponds to knowing the answer to a maximal number of
  questions (i.e., measurements described by one-dimensional orthogonal
  projectors).
\end{quotation}
The language about ``knowing the answer to a maximal number of
questions'' is redolent of ideas that QBism learned after much
scrutiny to leave behind.  These ideas go by names like ``the
eigenstate-eigenvalue link'' and ``the EPR criterion of reality'';
their underlying, unexamined premise is that a \emph{probability-1
  prediction} equates to an \emph{objective, agent-independent
  physical truth.}

How could we ever reconcile this language with the QBist insistence
that a quantum measurement does not simply read off a pre-existing
physical quantity?  It just doesn't work.  Every instance of ``maximal
information'' in CFS 2002 is an ironic echo of the traditional mantra
in a scientist's concluding remarks:\ When it came to the matter of
conceptual consistency, \emph{further research was needed.}

``Knowing the answer to a maximal number of questions'' applies to the
Spekkens toy model, by construction~\cite{Spekkens:2007}. In this
model, the fundamental atom is a system with four possible physical
states. The observer is restricted never to know more than one bit of
information about an atom that would require two bits to describe
fully. It follows that there are three possible binary-valued tests
that the observer can perform on an atom, but no state of knowledge
can allow the answer to more than one of them to be
foreseen. Consequently, the posit that ``Maximal information
\ldots\ corresponds to knowing the answer to a maximal number of
questions'' does not really get at anything uniquely quantum at
all~\cite{Spekkens:2007}.

QBists have argued that the restriction in certainty is not
fundamental, but rather derived. In order to explore this point, we
must do something that is doubtless anathema in philosphical
circles:\ discuss new technical developments rather than old
terminology. The crucial idea is the concept of a \emph{reference
  measurement}~\cite{Appleby:2017}.  Now that the kilogram is no
longer defined as a particular lump of platinum-iridium
alloy~\cite{Conover:2018}, there is room in the vault for a Bureau of
Standards quantum measurement device. Consider a physicist Alice who
has a qubit system in her possession. She can go to the Bureau of
Standards and drop her system into the standard measurement device for
qubits.  Such a device must have at least four possible outputs that
it can generate; that is, Alice's mathematical representation of it
must be a POVM with at least four elements. Let $p(H_i)$ denote
Alice's probability for obtaining outcome number $i$. Suppose that she
intends to perform some \emph{other} measurement $\{E_j\}$:\ perhaps a
von Neumann test corresponding to an orthonormal basis, or perhaps a
trine POVM~\cite{Fuchs:1997}, or even a ``noisy icosahedron'' POVM
relevant to the theory of exceptional Lie
algebras~\cite{Stacey:2019c}. Let $r(E_j|H_i)$ denote her probability
for eliciting outcome $j$ in this other experiment \emph{given} that
she has performed the Bureau of Standards reference POVM and obtained
outcome $i$ in it first.  Quantum theory then furnishes the tools for
Alice to compute her probability $q(E_j)$ for eliciting outcome $j$ in
this other measurement \emph{without} her carrying out the reference
experiment first.  The vector of these probabilities will be
\begin{equation}
  q = \mu(p, r),
\end{equation}
where $\mu$ is some function that depends upon the details of the
reference measurement.  A simple and illuminating choice is to make
the reference measurement a POVM that corresponds to a regular
tetrahedron inscribed in the Bloch sphere.  For example, letting $a$
and $b$ take the values $\pm 1$, then the four positive semidefinite
operators
\begin{equation}
  H_{ab} = \frac{1}{4}\left(
  I + \frac{1}{\sqrt{3}}(a\sigma_x + b\sigma_y + ab\sigma_z)
  \right)
\end{equation}
sum to the identity and thus constitute a POVM.  With $p(H_{ab}) =
\tr(\rho H_{ab})$ by the Born Rule and $r(E_j|H_{ab}) = 2\tr(E_j
H_{ab})$ by the L\"uders Rule, we have
\begin{equation}
  q(E_j) = \tr(\rho E_j) = \sum_{ab}\left[3p(H_{ab}) - \frac{1}{2}\right]
  r(E_j|H_{ab}).
\end{equation}
In this case, the function $\mu$ takes the form of the classical Law
of Total Probability but with an elementwise deformation of the
probability vector $p$.  The reference measurement establishes a
mapping from density matrices into probability vectors, thereby
yielding a wholly probabilistic representation of the quantum theory
of a qubit.  Not all probability vectors $p$ correspond to valid
quantum states in this representation.  In fact, with any
\emph{minimal informationally complete} (MIC) experiment as the
reference POVM, the state space is mapped into a proper subset of the
four-outcome probability simplex.  No more than \emph{one} entry in a
valid probability vector $p$ can be equal to zero.  Or, geometrically
speaking, the vertices of the reference probability simplex are
unavailable. This has the character of an uncertainty
principle:\ Alice's state of expectation can only be so sharp.  But
this is just a consequence of a deeper truth, namely that because
intrinsic ``hidden variables'' do not exist, one should not use the Law of
Total Probability to intermediate between different
experiments~\cite{DeBrota:2018b, Stacey:2019b}.

Later, CFS 2002 gets to this problematic passage:
\begin{quotation}
  \noindent In the classical case an i.i.d.\ assignment is often the
  starting point of a probabilistic argument.  Yet in Bayesian
  probability theory, an i.i.d.\ can never be strictly justified
  except in the case of maximal information, which in the classical
  case implies certainty and hence trivial probabilities. The reason
  is that the only way to be sure all the trials are identical in the
  classical case is to know everything about them, which implies that
  the results of all trials can be predicted with certainty
  [Jaynes~\cite{Jaynes:2003}].
\end{quotation}

Note that the citation supporting this argument is to E.\ T.\ Jaynes'
unfinished textbook~\cite{Jaynes:2003}. A QBist naturally asks, ``If
my probabilities really are \emph{mine,} then who's to stop me from
choosing an i.i.d.\ prior? Experience may lead me to revise my beliefs
away from that prior, but I have every right to assert it in the first
place.''  One could square the argument against the legitimacy of
classical i.i.d.\ priors with a Jaynesian view, but ultimately not
with a Ramseyan one. Saying ``to be sure all the trials are
identical'' amounts to saying ``to be sure the probabilities are
physically equal''. The argument in CFS 2002 is a relic of an
objective-Bayesian interpretation.  One way to express this shift of
interpretation is to say that in CFS 2002, probabilities are epistemic
(about knowledge), while in QBism proper, they are doxastic (about
belief).  CFS 2002 is saying that an i.i.d.\ prior is only justified
when the ratio of up to down probabilitons is constant across all
the trials, and that is just not a kind of Bayesianism that QBism can
endorse.

CFS 2002 declares (italics in original),
\begin{quotation}
  \noindent Since one of the chief challenges of Bayesianism is the
  search for methods to translate information into probability
  assignments, \emph{Gleason's theorem can be regarded as the greatest
    triumph of Bayesian reasoning.}
\end{quotation}
From a QBist perspective, this is peculiar. Gleason's theorem proves
that it is possible to chop off part of the standard formalism of
quantum theory and then re-grow it from the
remainder~\cite{Gleason:1957}. More specifically, Gleason showed that
if measurements correspond to orthonormal bases on a Hilbert space,
and if the probability of a measurement outcome does not depend upon
which basis the corresponding vector is embedded in, then any
consistent way of assigning probabilities to measurement outcomes has
to take the form of the Born rule. Thus, if $\Pi$ is a projection
operator and $p(\Pi)$ is the probability ascribed to obtaining the
outcome corresponding to $\Pi$, then we must have
\begin{equation}
  p(\Pi) = \tr(\rho \Pi)
\end{equation}
for some density matrix $\rho$. Both the set of valid $\rho$ and the
rule for what to do with a $\rho$ come tumbling out of Gleason's
insight. This is of course pertinent to the project of reconstructing
quantum theory, a task to which much QBist and QBist-adjacent effort
has been devoted --- but Gleason's theorem itself has barely figured
in that effort. Why? One reason is that the premises of Gleason's
theorem are themselves rather late in the game:\ Gleason's starting
point is a Hilbert space and orthonormal bases upon it. The natural
question is thus how to arrive at Hilbert space --- out of all the
mental contrivances that the mathematicians have conjured, why that
very particular class of structure? The stated goal of the
reconstruction project in which Fuchs and others have participated is
to derive complex Hilbert space, linear operators, the space of valid
quantum states and all the rest of the formalism from principles that
are more deeply rooted. In that light, Gleason's theorem is more a
proof of principle, a historically significant demonstration that the
machinery \emph{can} be taken apart and rebuilt, rather than ``the
greatest triumph'' of anything.

Busch~\cite{Busch:2003} --- and, later, independently Caves \emph{et
  al.}~\cite{Caves:2004} --- proved an analogue of Gleason's theorem
in which POVMs are the basic notion of measurement.\footnote{Busch
  discovered this result while salvaging von Neumann's attempt at a
  no-hidden-variables proof. The structure of POVMs allows one to make
  an additivity condition that replaces the unwarranted assumption in
  von Neumann's argument, the postulate that had been criticized by
  Hermann and probably by Einstein~\cite{Hermann:1935, Hermann:1935b,
    Wick:1995, Mermin:2018, Stacey:2019d}.}  Gleason's original
theorem fails for two-dimensional Hilbert spaces. The essential reason
is that in two dimensions, one cannot hold one vector of a basis in
place and twirl the rest of the basis around to generate multiple
distinct measurements. This difficulty does not apply to the POVM
version of the theorem. The question of which classes of measurements
allow the proof of a Gleason-type theorem continues to be
studied~\cite{Granstroem:2006, Wright:2019a, Wright:2019b}. For the
present purposes, we need only note that the emphasis on orthonormal
bases is another way that CFS 2002 does not read at all like the
genuinely QBist writings of Fuchs and Schack.

In my experience, Gleason's theorem is unfamiliar to many physicists,
and when they learn of it, they may find it unsatisfying. This
warrants a moment of consideration. I suspect that they want a story
about energy flow, thermalization, a phase transition manifesting as
symmetry breaking. (I went to physicist school too!) But here we have
no initial coarse approximation by dimensional analysis, no
semi-heuristic judgments about the comparative strengths of different
couplings. The state space and the Born Rule just fall out of the
geometry, like a regular pentagon from a nest of construction
lines. The emotional reaction is to say that there is no
\emph{physics} in it.\footnote{Zurek~\cite{Zurek:2018} calls Gleason's
  proof ``rather complicated''. True, but it's no Four Color Theorem,
  and a profession that boasts of using over 6300 tenth-order Feynman
  diagrams to compute the electron magnetic moment~\cite{Aoyama:2015}
  has no right to complain about anything that is merely ``rather
  complicated''. Zurek further says that Gleason ``provides no
  motivation why the measure he obtains should have any physical
  significance --- i.e., why should it be regarded as
  probability''. This exchanges cart and horse. The ethos of Gleason's
  theorem is to start with structures that do not \emph{look}
  probabilistic in character (Hilbert spaces and orthonormal bases, or
  orthomodular lattices if you're a quantum logician~\cite{Soler:1995,
    Holland:1995}). Then one says that to be meaningful, a physical
  theory must at least make statistical predictions. One introduces
  Gleason's ``frame functions'' \emph{because} one is motivated to
  find probabilities. And thus, all the structures that follow
  naturally inherit a probabilistic character, including the set of
  pure states. Of course, this clashes with the widespread gut-level
  conviction that a state vector $\ket{\psi}$ simply cannot be
  probabilistic in nature, and thus with Zurek's more specific claim
  that symmetries within a $\ket{\psi}$ are somehow
  ``objective''. Most arguments in quantum foundations are,
  ultimately, gastric distress.} But the math --- intricate,
laborious, much simplified by POVMs --- does work. The lesson, to a
certain mindset, is that pursuing a ``physics answer'' in the
pedestrian sense is redundant, needless, an effort better spent
elsewhere.

Next, we consider the interpretation that CFS 2002 places on the
\emph{quantum de Finetti theorem}~\cite{Hudson:1976, Caves:2002b,
  Fuchs:2004, Brandao:2017}.  This is a quantum analogue of the de
Finetti theorem in classical probability theory, which provides a
viable meaning for the term ``unknown probability'' in a subjectivist,
or personalist, form of Bayesianism.  Consider a scenario in which an
agent wishes to conduct a long experiment, made up of many successive
trials.  We can represent the outcome of each trial by a random
variable $x_j$, and Alice assigns a joint probability distribution
$p(x_1,x_2,\ldots,x_N)$ over the possible outcomes of an $N$-trial
experiment.  Imposing two conditions on this joint distribution turns
out to simplify its form dramatically.  First, we require that it be
\emph{finitely exchangeable}:\ Its value is invariant under
permutations of its arguments.  If $\pi$ is any permutation of the
indices $\{1,\ldots,N\}$, then
\begin{equation}
p(x_1,\ldots,x_N) = p(x_{\pi(1)},\ldots,x_{\pi(N)}).
\end{equation}
Second, we require that Alice's $p(x_1,\ldots,x_N)$ be
\emph{extendable,} in the following manner.  For any integer $M > 0$,
there must be a finitely exchangeable distribution with more arguments,
$p_{N+M}$, such that
\begin{equation}
p(x_1,\ldots,x_N)
 = \sum_{x_{N+1},\ldots,x_{N+M}}
    p_{N+M}(x_1,\ldots,x_N,x_{N+1},\ldots,x_{N+M}).
\end{equation}
These two requirements make precise the idea that Alice's probability
assignment $p$ derives from an arbitrarily long sequence of random
variables, the order of which is, in Alice's judgment,
inconsequential.  We say that a $p$ which satisfies both conditions,
finite exchangeability and extendable, is \emph{exchangeable.}

Let $\Delta_k$ denote the space of valid probability assignments over
$k$ outcomes. Then, the classical de Finetti theorem shows that
exchangeability implies that
\begin{equation}
p(x_1,\ldots,x_N)
 = \int_{\Delta_k} d\vec{p}\, P(\vec{p}\,)\, p_{x_1}\cdots p_{x_N}
 = \int_{\Delta_k} d\vec{p}\, P(\vec{p}\,)\, p_1^{n_1}\cdots p_k^{n_k},
\label{eq:de-finetti}
\end{equation}
where $P(\vec{p})$ is properly normalized over $\Delta_k$:
\begin{equation}
\int_{\Delta_k} d\vec{p}\, P(\vec{p}\,) = 1.
\end{equation}

The quantum version replaces an integral over the probability simplex
$\Delta_k$ with an integral over quantum state space, furnishing a
representation of exchangeable quantum-state ascriptions:
\begin{equation}
  \rho^{(N)} = \int d\rho\, P(\rho) \rho^{\otimes N}.
\end{equation}
Just as the classical de Finetti theorem revealed how the term
``unknown probability'' is merely a convenient shorthand, so does the
quantum de Finetti theorem for ``unknown quantum state''.

Having established this background, we turn to the following passage
from CFS 2002:
\begin{quotation}
  \noindent Exchangeability permits us to describe what is going on in
  quantum-state tomography. Suppose two scientists make different
  exchangeable state assignments and then jointly collect data from
  repeated measurements. Suppose further that the measurements are
  ``tomographically complete''; i.e., the measurement probabilities
  for any density operator are sufficient to determine that density
  operator. The two scientists can use the data $D$ from an initial
  set of measurements to update their state assignments for further
  systems. In the limit of a large number of initial measurements,
  they will come to agreement on a particular product state
  $\hat{\rho}_D \otimes \hat{\rho}_D \otimes \cdots$ for further
  systems, where $\hat{\rho}_D$ is determined by the data. \emph{This
    is what quantum-state tomography is all about.} The updating can
  be cast as an application of Bayes's rule to updating the generating
  function in light of the data [Schack, Brun and
    Caves~\cite{Schack:2001}]. The only requirement for ``coming to
  agreement'' is that both scientists should have allowed for the
  possibility of $\hat{\rho}_D$ by giving it nonzero support in their
  initial generating functions.
\end{quotation}
QBism refuses to go down this path. Indeed, it balks at the first
step. ``Suppose two scientists make different exchangeable state
assignments and then jointly collect data from repeated measurements''
--- no, we're stopping right there. QBism insists that
\emph{measurement outcomes are personal to the agent who elicits them.}

A QBist take on the quantum de Finetti theorem puts all the state
assignments into a single user's internal mesh of beliefs. Alice
supposes that tomorrow, she will make a multipartite, exchangeable
state ascription, perhaps $\rho_1^{(N)}$ or perhaps
$\rho_2^{(N)}$. Using the quantum de Finetti theorem, she represents
the first joint state as a ``meta-probability density'' $P_1(\rho)$:
\begin{equation}
  \rho_1^{(N)} = \int d\rho\, P_1(\rho) \rho^{\otimes N},
\end{equation}
and likewise for $\rho_2^{(N)}$. If the density functions $P_1(\rho)$
and $P_2(\rho)$ have at least a little agreement, Alice can
\emph{expect} that her initial choice will wash out. That is, her
expectation \emph{now} for her \emph{future mesh of beliefs} is that
the choice between the ascriptions $P_1(\rho)$ and $P_2(\rho)$ will
eventually become inconsequential. So, there is definitely a story to
be told about the mathematics, perhaps a rather important one, but it
is not the story given in CFS 2002.\footnote{For a pedagogical
  introduction to the ``probabilities for future probabilities''
  thinking, see~\cite[\S 5.1]{Stacey:2015}. For a technical result
  motivated by ``expected changes in expectation'' concerns,
  see~\cite{DeBrota:2019}.}

Recalling Fuchs' three distinguishing characteristics of QBism, it is
difficult to argue that they are present in CFS 2002. The ``formal
structure'' is not reshaped in look or in feel, merely propped up by a
couple extant theorems, one of them known since 1957. Two levels of
radical personalism are not present --- there is only one, and it is
but tepidly embraced, while the other is flatly contradicted. The
third characteristic, the ``start of a great journey'', receives an
endorsement of sorts in the paper's send-off, but one so disconnected
from everything that went before, it reads more like an afterthought
than an outlook.

\section{Caves--Fuchs--Schack (2007)}

Caves, Fuchs and Schack moved away from some positions held in their
2002 paper just a few years later. ``Subjective probability and
quantum certainty'' (\cite{CFS:2007}, hereinafter CFS 2007) is the
last work coauthored by all three of them, since as that paper was
being written and published, it became clear that Caves disagreed with
Fuchs and Schack on various points that are essential to Fuchs and
Schack's further development of QBism.\footnote{Caves has confirmed to
  me (e-mail, 24 November 2019) that he does not subscribe to QBism.}  In
their introduction, they state the following:
\begin{quotation}
  \noindent In a previous publication [CFS 2002], the authors were
  confused about the status of certainty and pure-state assignments in
  quantum mechanics and thus made statements about state preparation
  that we would now regard as misleading or even wrong.
\end{quotation}
Discussions leading up to this change of heart can be found
in~\cite[pp.\ 193 ff.]{Fuchs:2014}.

There is less material in CFS 2007 that is overtly contra-QBist than
there is in CFS 2002, yet Fuchs' later warning covers it as
well~\cite{Fuchs:2010}:
\begin{quotation}
  \noindent The present work, however, goes far beyond those
  statements in the metaphysical conclusions it draws---so much so
  that the author cannot comfortably attribute the thoughts herein to
  the triumvirate as a whole. Thus, the term QBism to mark some
  distinction from the known common ground of Quantum Bayesianism.
\end{quotation}
One would think that this hazard sign would prompt a degree of caution
to be taken before treating all the citations in a list of ``Quantum
Bayesian'' papers on equal footing.\footnote{Indeed, such a list of
  predecessors occurs in Fuchs and Schack's \booktitle{Reviews of
    Modern Physics} article~\cite{Fuchs:2009}. Cursory inspection
  reveals that the views tallied there should not be identified with
  QBism or with each other.  For example, the list includes Appleby's
  ``Facts, Values and Quanta''~\cite{Appleby:2005}, which takes pains
  to remain distinct from the other Quantum Bayesian writings of that
  period:\ ``I should say that I do not entirely agree with them about
  that. [\ldots\!] My feeling is that a completely satisfactory
  theoretical account has yet to be formulated.'' And so forth.} We
can see a trace of a divergence already manifesting in this aside:
\begin{quotation}
  \noindent Bayesian updating is consistent, as it should be, with
  logical deduction of facts from other facts, as when the observed
  data $d$ logically imply a particular hypothesis $h_0$, i.e., when
  ${\rm Pr}(d|h) = 0$ for $h \neq h_0$, thus making ${\rm Pr}(h_0|d) =
  1$. Since the authors disagree on the implications of this
  consistency, it is fortunate that it is irrelevant to the point of
  this paper. That point concerns the status of quantum measurement
  outcomes and their probabilities, and quantum measurement outcomes
  are not related by logical implication.
\end{quotation}

CFS 2007 is evasive on the question of whether measurement outcomes
are personal to the agent who elicits them. They write of ``the facts
an agent acquires about the preparation procedure'', and they say,
``The occurrence or nonoccurrence of an event is a \emph{fact} for the
agent.'' But if the occurrence or nonoccurrence of an event is a fact
for \emph{everybody,} it is a fact for a specific agent too. Without
forthrightly clarifying this point, CFS 2007 does not qualify as
QBist. Indeed, CFS 2007 rather undermines the point, with loose talk
about ``two agents starting from the same facts, but different
priors'' and the like. A QBist description would instead involve a
single agent, considering the same set of quantitative data points
with either of two background meshes of belief (compare
\cite{Fuchs:2009b, Stacey:2016b}).

Before moving on, we note that CFS 2007 briefly discusses the Lewisian
``objective chance'' philosophy. Fuchs would shortly thereafter find
this discussion weak enough to be irrelevant, or potentially even
counterproductive~\cite[p.\ 1287]{Fuchs:2014}. Overall, CFS 2007 is
the product of too many compromises among its three authors to fairly
reflect the view of any of them.

\section{The First Samizdat}
\booktitle{Notes on a Paulian Idea} (2003), later reissued as
\booktitle{Coming of Age with Quantum Information} (2011), is an
edited collection of e-mail correspondence that Fuchs made public to
``back up the hard drive'' after the Cerro Grande fire destroyed his
family's home in Los Alamos~\cite{Fuchs:2003}. My colleagues and I
have elsewhere~\cite{Appleby:2017} quoted passages from this document
to show how later research has fulfilled their aspirations almost to
the letter. Here, I take the opposite tack.

One theme that is quite surprising to a reader familiar with QBism
proper is how frequently Fuchs insists upon multiple agents being
necessary to reveal the hidden ontological lesson of quantum
theory. This stands in stark contrast to the
``I-I-me-me-mine!''\ declaration of Fuchs' later
manifesto~\cite{Fuchs:2010}. In addition, the stories of multiple
agents intermingle with the theme that the deep physical principle of
quantum theory is captured by information-disturbance tradeoffs. This
idea had vanished by the time QBism proper was articulated. We can
point to multiple reasons why it could easily have fallen by the
wayside:\ The quantitative expressions of it started out dense and not
too illuminating~\cite{Fuchs:1996, Fuchs:1998}, and despite some
promising indications~\cite{Fuchs:1998b}, they never really
simplified.\footnote{Asks Fuchs in~\cite{Fuchs:1998b}, ``Why is the
  world so constituted that binary preparations can be put together in
  a way that the whole is more than a sum of the parts, but never more
  so than by $Q \approx 0.202$ bits?'' Note that the bound of $Q
  \approx 0.202$ bits is attained when the two alphabet states are
  drawn from two different MUB; while by another measure of
  ``quantumness'' in that paper, the average global fidelity, two
  qubit states are ``most quantum with respect to each other'' when
  they are drawn from a SIC.} Moreover, the basic phenomenon of ``no
information gain without disturbance'' ended up being too easy to
reproduce in theories with underlying local-hidden-variable
models.\footnote{True, one might be able to establish a difference in
  the precise shape of the tradeoff curves (compare
  \cite{Spekkens:2018}), but that is too slender a reed to hang the
  distinction between quantum and classical upon.  Like NASA, at the
  critical juncture we want a declaration that is \emph{go/no-go.}}

In an 18 September 1996 letter to David Mermin, we find an early
occurrence of the slogan that Fuchs and Schack would later spurn:
\begin{quotation}
  \noindent One always assigns probabilities based on incomplete
  information; it is just that in quantum physics ``maximal
  information is not complete.''
\end{quotation}
We find this again in another letter to Mermin, this one dated 23 July 2000:
\begin{quotation}
  \noindent The theory prescribes that no matter how much we know
  about a quantum system---even when we have \emph{maximal}
  information about it---there will always be a statistical
  residue. There will always be questions that we can ask of a system
  for which we cannot predict the outcomes. \emph{In quantum theory,
    maximal information is not complete and cannot be completed.}
\end{quotation}
Even sentences that were blessed with italics can turn out to be
quite wrong-headed.

We can uncover early intimations of the QBist desire to find in
quantum theory an ontological lesson, without na{\"\i}vely identifying
the elements in the mathematical formalism with an ontology. However,
the place and manner of the search is not yet QBist.  From a 4 January
1998 letter to Greg Comer:
\begin{quotation}
  \noindent The ``fact'' that \emph{my} information-gathering yields a
  disturbance to \emph{your} predictions is the only ``physical'' (or
  ontological) statement that the theory makes; all the rest of the
  structure is ``law of thought'' subject to that consideration. To
  put it another way, quantum theory is a theory of ``what we have the
  right to say'' in a world where the observer cannot be detached from
  what he observes. It is that and nothing more.
\end{quotation}
And again on 22 April 1999:
\begin{quotation}
  \noindent Our experimentation on the world is not without
  consequence. When \emph{I} learn something about an object,
  \emph{you} are forced to revise (toward the direction of more
  ignorance) what you could have said of it.
\end{quotation}
Likewise, in a 30 August 1998 letter to Adrian Kent:
\begin{quotation}
  \noindent Suppose in a few years I could come up with a clear, precise
  statement of what it is that I'm trying to get at. Something in
  essence that takes away the vagaries of the statement: ``The world
  in which we live happens to have a funny property. It is that
  \emph{my} information gathering about something you know, causes
  \emph{you} to lose some of that knowledge \ldots\ and this happens
  even in the case that you know all my actions precisely. Physical
  theory, and quantum mechanics in particular, is about what we can
  say to each other and what we can predict of each other in spite of
  that funny property.'' Would that constitute something that fulfills
  your Desideratum \#1?
\end{quotation}
More concisely, in a letter on 2 September 1998:
\begin{quotation}
  \noindent Disturbance to what? To each other's descriptions, nothing
  more.
\end{quotation}

The themes of information tradeoffs and multiplicity of agents are
developed rather extensively in correspondence with David Mermin. On 8
September 1998:
\begin{quotation}
  \noindent It is crucial to my point of view that there be at least
  \emph{two} players and \emph{two} quantum states in the
  game. [\ldots\!] I think one of the troubles in our founding
  fathers' discussions is that they continually focused their
  attention on \emph{one} observer making measurements on a quantum
  system described by \emph{one} (known) quantum state. This led them
  to say things---in language similar to some of the specimens in your
  note---like, ``The gain of knowledge by means of an observation has
  as a necessary and natural consequence the loss of some other
  knowledge.'' (Pauli) Without at least a second player in the game,
  those gains and losses hardly seem to be sensible concepts to me:
  they can only refer to the observer's attempt to ascribe one or
  another classical picture to the quantum system in front of him.
  Since we know---from Bell's argument and the religion of
  locality---that it is not reasonable to assume that those classical
  variables (correlata) are there and existent without our prodding,
  it is hard to call the revelation of a measurement outcome a ``gain
  of knowledge.'' What did you learn about the world that was there
  before your looking? Nothing. However, throw a second player into
  the game and that situation changes. Those random quantum outcomes
  now have something \emph{existent,} some unknown truth, that they
  can be correlated with. The revelation of an outcome really can
  correspond to a ``gain of knowledge,'' but you need at least two
  information processing units in the world for that to be the case.
\end{quotation}
And, shortly thereafter in the same note,
\begin{quotation}
  \noindent In your original Ithaca paper you speak of the minimal
  requirements for a quantum mechanical universe: it is, you say, two
  qubits---two things to have correlation without correlata. I,
  however, am more afraid to go that far, i.e., to some
  final/overarching ontological statement. Instead, the most I think
  I'm willing to ask is, ``What are the minimal requirements for a
  physical \emph{theory}?'' And there, I think the answer is two
  ``theory makers'' and a physical system.
\end{quotation}
More dramatically still, on 20 July 2000:
\begin{quotation}
  \noindent Somehow I feel that I had an epiphany in Mykonos. Do you
  remember the parable of ``Genesis and the Quantum'' from my
  Montr\'eal problem set? And do you remember my slide of an empty
  black box with two overlays. The first overlay was of a big
  $\ket{\psi}$ (hand drawn in blue ink of course). I put the slide of
  the box up first, and said, ``This is a quantum system; it's what's
  there in the world independent of us.'' Then I put the first overlay
  on it and said, ``And this symbol stands for nothing more than
  [what] we know of it. Take us away and the symbol goes away too.'' I
  then removed the $\ket{\psi}$. ``But that doesn't mean that the
  system, this black box, goes away.'' Finally I put back up the
  $\ket{\psi}$ over the box, and the final overlay. This one says:
  ``Information/knowledge about what? The consequences of our
  experimental interventions into the course of Nature.''

  Well, now I've made another overlay for my black box slide. At the
  top it asks, ``So what is real about a quantum system?'' In the
  center, so that it ends up actually in the box, is a very stylistic
  version of the word ``Zing!'' And at the bottom it answers, ``The
  locus of all information-disturbance tradeoff curves for the
  system.'' In words, I (plan to) say, ``It is that zing of the
  system, that sensitivity to the touch, that keeps us from ever
  saying more than $\ket{\psi}$ of it. This is the thing that is real
  about the system. It is our task to give better expression to that
  idea, and start to appreciate the doors it opens for us to shape and
  manipulate the world.'' What is it that makes quantum cryptography
  go? Very explicitly, the zing in the system. What is that makes
  quantum computing go? The zing in its components!

  Anyway, I'm quite taken by this idea that's getting so close to
  being a technical one---i.e., well formed enough that one might
  check whether there is something to it. What is real of the system
  is the locus of information-disturbance (perhaps it would be better
  to say ``information-information'') tradeoff curves. The thing to do
  now is to show that Hilbert space comes about as a compact
  description of that collection, and that it's not the other way
  around. As I've preached to you for over two years now, this idea
  (though it was in less refined form before now) strikes me as a
  purely ontological one \ldots\ even though it takes inserting an
  Alice, Bob, and Eve into the picture to give it adequate
  expression. That is, it takes a little epistemology before we can
  get to an ontological statement.
\end{quotation}
The number of agents is getting out of hand --- not just an Alice,
with whom a QBist narrative would content itself, but a Bob and now an
Eve.  This is quite the dramatic excess in the light of QBism's
single-user focus (a later development codifed in response to drilling
down on the issue of Wigner's Friend~\cite[p.\ xli]{Fuchs:2014}).
\begin{flushright}
  ``What, four? thou saidst but two even now.''\\
  ``Four, Hal; I told thee four.''\\
  \medskip
  --- \booktitle{1 Henry IV, 2.4}
\end{flushright}

We do see a backing-away from some early choices of terminology, in a
1 July 2000 letter to Hideo Mabuchi:
\begin{quotation}
  \noindent I'm in Greece right now, just finished with the NATO
  meeting. Tomorrow morning I leave for Capri (the QCMC conference),
  and then finally join Kiki in Munich at the end of the week. My talk
  was pretty successful in Mykonos; I was pretty happy with it. For
  Capri I'm going to make a completely new one, this time based on the
  stuff I did with Kurt Jacobs. I've decided the best way to say what
  I've been hoping to get at. Question: ``If the wavefunction isn't
  real, then what is it that IS real about a quantum system?'' Answer:
  ``The locus of all information-information tradeoff curves that one
  can draw for such a system.'' (I've decided to stop calling it
  information-disturbance because it conveys bad imagery and
  preconceptions. The disturbance is to information, so why not just
  make it explicit.)
\end{quotation}
In retrospect, this prefigures Fuchs and Schack's later
excommunication of the ``maximal information is not complete'' slogan
--- but only in retrospect. It is still locked into an epistemic mindset,
rather than a doxastic one.

In an 11 December 2000 letter to Joseph Renes:
\begin{quotation}
  \noindent Why am I so obsessed with always having two players in the
  game? Because I want to connect all the concerns in quantum
  mechanics with Bayesianism as much as I can.
\end{quotation}
At this point in time, Fuchs pretty explicitly takes the convergence
among agents as the meat of Bayesianism, rather than making the
fundamental point the normative principle of consistency within a
single agent's mesh of beliefs, with inter-agent agreement a secondary
notion (when it can meaningfully be defined at all).

\section{Fuchs (2002)}

Having grounded ourselves in Fuchs' less formal solo-author writings
from the late 1990s, we are now in a better position to examine
``Quantum Mechanics as Quantum Information (and only a little
more)''~\cite{Fuchs:2002}. I have deferred discussion of this essay
until now, because much of it is technical development, and the
conceptual passages where it comes off most strongly non-QBist are
best appreciated after becoming familiar with the \booktitle{Notes on
  a Paulian Idea} era. Of the three distinguishing features of QBism,
we can discern the ``reliance on the mathematical tools of quantum
information theory to reshape the look and feel of quantum theory's
formal structure'', at least in a preliminary way. And, not bound by
the length and style constraints of a journal article's
``Conclusions'' section, Fuchs takes the opportunity to press the
``start of a great journey'' theme. But there is still only one level
of personalism:\ Probabilities are personal, but \emph{experiences}
are not. A brief excerpt suffices to show that Fuchs attempts to
launch the ``great journey'' just as he did at Mykonos:
\begin{quotation}
  \noindent The wedge that drives a distinction between Bayesian
  probability theory in general and quantum mechanics in particular is
  perhaps nothing more than this ``Zing!''\ of a quantum system that
  is manifested when an agent interacts with it. It is this wild
  sensitivity to the touch that keeps our information and beliefs from
  ever coming into too great of an alignment. The most our beliefs
  about the potential consequences of our interventions on a system
  can come into alignment is captured by the mathematical structure of
  a pure quantum state $\ket{\psi}$. Take all possible
  information-disturbance curves for a quantum system, tie them into a
  bundle, and \emph{that} is the long-awaited property, the input we
  have been looking for from nature. Or, at least, that is the
  speculation.
\end{quotation}
The \emph{first} sentence would read fine coming from a QBist, but the
rest goes barrelling down a blind alley.

The technical discussions hold up rather better, and the paper is
noteworthy as an early example of the MIC-as-reference-measurement
idea. The particular class of MIC it discusses, later designated the
\emph{orthocross MICs,} still has some open conjectures about
it~\cite{DeBrota:2018c}.

Fuchs gives an argument for why the tensor product rule for composing
state spaces follows from Einstein locality and a Gleason-type
context-independence condition. This proof may be of significance to a
category theorist~\cite[\S 2.3]{Riehl:2016}, as it deduces the tensor
product from the requirement that the functions of interest be linear
on both halves of the composite system.  However, it does go somewhat
against the grain of later reconstruction work with Schack and
others. Those efforts focused on deriving the state and measurement
spaces of a single system, which can then be resolved into components
if desired. In other words, the emphasis shifted from
\emph{composition} to \emph{decomposition.}

The 2002 paper leaves open the question of why the joint states for a
bipartite system should be specifically \emph{positive semidefinite}
operators on the tensor-product space. The later literature provides
at least one answer to this question~\cite{Barnum:2010,
  DeLaTorre:2012}, but at the cost of assumptions that may feel
unsatisfying on account of being physically under-motivated or
mathematically over-powered. (For example, why in the grand scheme of
things should the set of entangled pure states form a continuum?)
There may yet be a theorem or two worth proving here.

Examining the motivations interleaved between the equations, we find
another conceptual issue that marks the 2002 paper as not yet QBist.
It is the distinction between doxastic \emph{consistency conditions} and
\emph{update rules,} a point that Fuchs and Schack did not fully
resolve until the better part of a decade
later~\cite{Fuchs:2012}. Quoting~\cite{DeBrota:2018},
\begin{quotation}
\noindent Adopting a personalist Bayesian interpretation of
probability does \emph{not} mean treating all changes of belief as
applications of the Bayes rule.  This is shocking to some people!  And
distancing ourselves from the dogmatists who claim to follow that
creed is one reason why we prefer \emph{QBism} over ``Quantum
Bayesianism''.

In the tradition of Ramsey, Savage and de Finetti, there are
consistency conditions that an agent's probability assignments should
meet at any given time, \emph{and then} there are guidelines for
\emph{updating} probability assignments in response to new
experiences.  Going from the former to the latter requires making
extra assumptions --- the two are not as strongly coupled as many
people think. The Bayes rule is not a condition on how an agent
\emph{must} change her probabilities, but rather a condition for how
she should \emph{expect} that she will modify her beliefs in the light
of possible new experiences. For this observation, we credit Hacking,
Jeffrey and van Fraassen.
\end{quotation}
Fuchs' writing in 2002 had not yet distinguished the crucial gap
between a rule for how Alice \emph{must} change her beliefs and a
criterion for how Alice should \emph{expect} today that she will act
tomorrow.

This is a slip-up we encounter now and then in conversations with
people who have only heard a little about QBism, usually
secondhand. (Other confusions --- ``But how does QBism explain X?''
--- typically occur when the interlocutor has unwittingy switched
from subjective probability to objective, or from a first-person
perspective to third-person, midway through a thought process. These
are habits which take discipline to avoid, at least at first.)
Evading this mental trap is another good reason not to take Fuchs'
2002 salvo as a definitive, genuinely QBist position statement.

\section{Fuchs--Peres (2000)}

On occasion, we have seen CFS 2002 cited on its own to define QBism
(for example, in~\cite{Vaidman:2019, Laudisa:2019}). A similar yet
more egregious misattribution occurs in an article by
Jaeger~\cite{Jaeger:2019}, which equates QBism with the 2000
\booktitle{Physics Today} piece coauthored by Fuchs and Peres,
``Quantum Theory Needs No `Interpretation'\,''~\cite{Fuchs:2000}.  I
must regretfully report that Asher Peres was no QBist.

The specific point at which Jaeger elides the difference between QBism
and \emph{fin-de-si\`ecle} Fuchs--Peres is the following, which
attributes an opinion of the latter to the former:
\begin{quotation}
  \noindent One QBist claim is that ``quantum theory does \emph{not}
  describe physical reality. What it does is provide an algorithm for
  computing \emph{probabilities} for the macroscopic events (`detector
  clicks') that are the consequence of our experimental
  interventions. This strict definition of the scope of quantum theory
  is the only interpretation ever needed, whether by experimenters or
  theorists.''
\end{quotation}
But this is not a QBist claim. Of course, it predates QBism
chronologically, but also, it contradicts what QBism actually stands
for, and in a series of rather blunt and obvious ways.  It is helpful
here to quote a letter from Fuchs' second
samizdat~\cite[p.\ 1011]{Fuchs:2014}, sent on 19 June 2005 to Greg
Comer:
\begin{quotation}
  \noindent First off, I wish I had never said, ``quantum theory does
  not describe physical reality''---I really only meant ``the wave
  function does not describe reality'' and should have stuck with that
  formulation.  But more importantly, what precisely are these
  ``consequences of our interventions''?  From the wording we used,
  one surely gets the impression that, whatever they are---we said
  ``detector clicks,'' but what a glib phrase!---they somehow live
  outside of the agent performing the experiment.  And I guess that's
  what I thought at the time.
\end{quotation}
So, ``detector clicks'' is misleading. Moreover, ``macroscopic'' is a
red herring, a relic of earlier generations' shifty grasp on what
might differentiate quantum from classical. For example, an agent
whose species has evolved eyes just a bit better than human ones might
have seen individual photons flashing on a cold and lonely night. Such
an agent might regard the direct personal experience of a single
photon as a microscopic event, but they can employ the ``user's
manual'' that is quantum theory just as well as humans do. In brief,
the micro/macro distinction is not, to a QBist, fundamental.

Would a QBist agree that a ``strict definition of the scope of quantum
theory is the only interpretation ever needed, whether by
experimenters or theorists''? No, \emph{ever needed} is all wrong
there.  Nothing that is all that's \emph{ever needed} can be the start
of a great adventure.

The Fuchs--Peres collaboration has a very un-QBist reliance
upon the first-person plural. As David Mermin has
noted~\cite{Mermin:2018b},
\begin{quotation}
  \noindent There is a little remarked upon but important ambiguity in
  the first person plural. When Heisenberg says that quantum states
  are about \emph{our} knowledge, ``our'' can mean all of us
  collectively or it can mean each of us individually. [\ldots\!]  To
  avoid ambiguity it is better to say ``My (your, Alice's) quantum
  state assignments encapsulate my (your, her) belief'' to avoid
  misreadings based on implicit assumptions of a unique state
  assignment or of common knowledge.
\end{quotation}
The Fuchs--Peres essay has some affinities with the Rudolf Peierls
opinion piece from a decade earlier~\cite{Peierls:1991} that they
discussed during the writing process~\cite{Fuchs:2003}.  Mermin
observes that Peierls would sound more QBist than perhaps any other
figure from the early generations of quantum physicists, if he had not
used the first-person plural collectively.  Instead, he propagated the
old confusions.  This applies with equal force to the Fuchs--Peres
collaboration.

Having dismissed ``our'', ``macroscopic'', ``detector clicks'', ``ever
needed'' and ``does not describe physical reality'', what about
``algorithm''? This, too, reflects an understanding that had not yet
matured. An algorithm is a step-by-step procedure that can be executed
mechanically.\footnote{At least, so it was in the year 2000. Nowadays,
  rather than the Knuthian sense of a procedure for a machine of known
  architecture, published so its performance can be analyzed, an
  algorithm is a trade secret that runs ``in the cloud'' and whose
  goal is to disguise injustice and inequality as objective
  logic~\cite{ONeil:2016}. O tempora, o mores.} For example, taking
the trace of the product of two matrices is a task for which an
algorithm might be written. In that sense, computing a
quantum-mechanical probability is algorithmic --- but there is a
deeper level of meaning, too, that the word \emph{algorithm}
misses. In QBism, a state vector $\ket{\psi}$ is not more
ontologically fundamental than, say, the probability of getting a
``$+$'' outcome in a spin-$z$ experiment. True, quantum theory
provides a rule for calculating $p_z(+)$ given a $\ket{\psi}$, but
when in life is one ever given a $\ket{\psi}$, other than the first
line of a textbook problem demanding that the student ``assume the
state vector is $\ket{\psi}$''?

The deeper truth is that quantum theory provides a \emph{normative
  lesson}: When Alice contemplates two or more von Neumann
measurements upon a system, she \emph{should strive} to make her
expectations for those different, mutually exclusive scenarios all
consistent with the Born Rule.  But if she detects an inconsistency
within her mesh of beliefs --- if she finds that there is no density
operator $\rho$ with which her varied probabilities are all in accord
--- the quantum formalism itself \emph{provides no algorithm} to
resolve that awkwardness~\cite{Fuchs:2017b}.

The novice at any art often begins by following a procedure --- say,
the exact volume measurements and timings given by \booktitle{The Joy
  of Cooking,} or the rubric for cranking through questions on the AP
Physics exam.  With further experience, one learns how to season for
taste, what can be substituted for chicken or for eggs, how to
linearize around the fixed points and so on.  The procedures are
always there to be relied upon when required --- chopping a root or
solving by radicals --- but they are not the soul of the matter.  So,
too, for quantum theory: Algorithms are what we use, not the sum total
of what we need.

Fuchs and Peres present a version of the Wigner's Friend
thought-experiment.  In their portrayal, Erwin applies quantum
mechanics to his colleague Cathy, who in turn applies it to a piece of
cake.  The Fuchs--Peres discussion is, from a QBist standpoint, rather
unforgivably sloppy about the distinctions between ontic degrees of
freedom, epistemic statements about ontic quantities and doxastic
statements regarding future personal experiences.

I would not have imagined it possible to declare that the Fuchs--Peres
opinion piece defines QBism, had I not seen it done in print.

\section{Conclusions}

Basically nothing posted on the arXiv before 2009 should be cited as
an example of QBism, no matter who the authors are.  All the older
writings fail in one or another readily apparent way to recognize at
least one point that later investigation found to be necessary for a
self-consistent interpretation of quantum mechanics.  That said,
various technical matters raised in those pre-QBist papers continue to
be interesting even though the metaphysical frontier has left their
original motivations far behind.

\bigskip

For conversations and correspondence, I thank Gabriela Barreto Lemos,
Carlton Caves, John B.\ DeBrota, Christopher Fuchs, Jacques Pienaar
and R\"udiger Schack. For a colloquium talk that inspired a technical
question I should be working on instead~\cite[\S VII]{Stacey:2019b}, I
thank Nicole Yunger Halpern. This research was supported by the John
Templeton Foundation. The opinions expressed in this publication are
those of the author and do not necessarily reflect the views of the
John Templeton Foundation.

\vfill
\begin{flushright}
  You may already know what a blow to the ego\\
  it can be to have to read over anything you\\
  wrote twenty years ago, even cancelled checks.\\
\medskip
--- Thomas Pynchon\\
\bigskip
Looking back it is clear, but so\\
much prevents you from seeing it.\\
\medskip
--- \href{https://blogs.scientificamerican.com/roots-of-unity/q-a-with-autumn-kent/}{Autumn Kent}

\end{flushright}

\begin{thebibliography}{999}
\bibitem{DeBrota:2018} J.\ B.\ DeBrota and B.\ C.\ Stacey,
  ``FAQBism,'' \arxiv{1810.13401} (2018).
  
\bibitem{Fuchs:2009} C.\ A.\ Fuchs and R.\ Schack, ``Quantum-Bayesian
  Coherence,'' \arxiv{0906.2187} (2009). A
  \hrefdoi{10.1103/RevModPhys.85.1693}{condensed version} was later
  printed as \booktitle{Reviews of Modern Physics} \textbf{85} (2013),
  1693.

\bibitem{Fuchs:2014} C.\ A.\ Fuchs, \booktitle{My Struggles with the
    Block Universe} (2014). Edited by B.\ C.\ Stacey, with a foreword
  by M.\ Schlosshauer. \arxiv{1405.2390}.

\bibitem{CFS:2002} C.\ M.\ Caves, C.\ A.\ Fuchs and R.\ Schack,
  ``\hrefdoi{10.1103.PhysRevA.65.022305}{Quantum probabilities as
  Bayesian probabilities},'' \booktitle{Physical Review A} \textbf{65}
  (2002), 022305, \arxiv{quant-ph/0106133}.

\bibitem{Pitowsky:2002} I.\ Pitowsky, ``Betting on the outcomes of
  measurements:\ A Bayesian theory of quantum probability,''
  \arxiv{quant-ph/0208121} (2002).
  
\bibitem{Bub:2009} J.\ Bub and I.\ Pitowsky, ``Two dogmas about
  quantum mechanics.'' In \booktitle{Many Worlds?:\ Everett, Quantum
    Theory, \& Reality} (Oxford University Press, 2010),
  \arxiv{0712.4258}.

\bibitem{Bub:2019} J.\ Bub, ``\,`Two dogmas' redux,''
  \arxiv{1907.06240} (2019).
  
\bibitem{Youssef:1994} S.\ Youssef,
  ``\hrefdoi{10.1142/S0217732394002422}{Quantum mechanics as Bayesian
  complex probability theory},'' \booktitle{Modern Physics Letters A}
  \textbf{9} (1994), 2571--2586, \arxiv{hep-th/9307019}.
  
\bibitem{Baez:2003} J.\ C.\ Baez, ``Bayesian Probability Theory and
  Quantum Mechanics,'' \url{http://math.ucr.edu/home/baez/bayes.html}
  (2003).

\bibitem{Leifer:2013} M.\ S.\ Leifer and R.\ W.\ Spekkens,
  ``\hrefdoi{10.1103/PhysRevA.88.052130}{Towards a formulation of
  quantum theory as a causally neutral theory of Bayesian
  inference},'' \booktitle{Physical Review A} \textbf{88} (2013),
  052130, \arxiv{1107.5849}.
  
\bibitem{Caticha:2007} A.\ Caticha, ``\hrefdoi{10.1063/1.2713447}{From
  objective amplitudes to Bayesian probabilities},'' \booktitle{AIP
  Conference Proceedings} \textbf{889} (2007), 62,
  \arxiv{quant-ph/0610076}.

\bibitem{Good:1983} I.\ J.\ Good, ``46,656 varieties of Bayesianism.''
  In \booktitle{Good Thinking:\ The Foundations of Probability and its
    Applications} (University of Minnesota Press, 1983).

\bibitem{CFS:2007} C.\ M.\ Caves, C.\ A.\ Fuchs and R.\ Schack,
  ``\hrefdoi{10.1016/j.shpsb.2006.10.007}{Subjective probability and
  quantum certainty},'' \booktitle{Studies in the History and
  Philosophy of Modern Physics} \textbf{38} (2007), 255--74,
  \arxiv{quant-ph/0608190}.

\bibitem{Fuchs:2013c} C.\ A.\ Fuchs, N.\ D.\ Mermin and R.\ Schack,
  ``\hrefdoi{10.1119/.1.4874855}{An introduction to QBism with an
  application to the locality of quantum mechanics},''
  \booktitle{American Journal of Physics} \textbf{82} (2014), 749--54,
  \arxiv{1311.5253}.
  
\bibitem{Fuchs:2017} C.\ A.\ Fuchs, ``On participatory realism.'' In
  \booktitle{Information and Interaction:\ Eddington, Wheeler, and the
    Limits of Knowledge}\ (Springer, 2017), \arxiv{1601.04360}.

\bibitem{Fuchs:2017b} C.\ A.\ Fuchs, ``Notwithstanding Bohr, the
  reasons for QBism,'' \booktitle{Mind and Matter} \textbf{15} (2017),
  245--300, \arxiv{1705.03483}.
  
\bibitem{Fuchs:2019} C.\ A.\ Fuchs and B.\ C.\ Stacey,
  ``QBism:\ Quantum Theory as a Hero's Handbook.'' In
  \booktitle{Proceedings of the International School of Physics
    ``Enrico Fermi,'' Course 197 -- Foundations of Quantum Physics},
  edited by E.\ M.\ Rasel, W.\ P.\ Schleich and S.\ W\"olk
  (Italian Physical Society, 2019). \arxiv{1612.07308}.
  
\bibitem{Cabello:2017}  A.\ Cabello, ``Interpretations of
  quantum theory: A map of madness.'' In \booktitle{What is Quantum
  Information?}\ (Cambridge University Press, 2017), \arxiv{1509.04711}.
  
\bibitem{Frauchiger:2018} D.\ Frauchiger and R.\ Renner,
  ``\hrefdoi{10.1038/s41467-018-05739-8}{Quantum theory cannot
  consistently describe the use of itself},'' \booktitle{Nature
  Communications} \textbf{9} (2018), 3711, \arxiv{1604.07422}.

\bibitem{Koberinski:2018} A.\ Koberinski and M.\ P.\ M\"uller,
  ``Quantum theory as a principle theory:\ insights from an
  information-theoretic reconstruction.'' In \booktitle{Physical
    Perspectives on Computation, Computational Perspectives on
    Physics} (Cambridge University Press, 2018), \arxiv{1707.05602}.

\bibitem{Schaffer:2019} K.\ Schaffer and G.\ Barreto Lemos,
  ``\hrefdoi{10.1007/s10699-019-09608-5}{Obliterating thinginess:\ An
  introduction to the ``what'' and ``so what'' of quantum physics},''
  \booktitle{Foundations of Science} (2019).
  
\bibitem{VonBaeyer:2016} H.\ C.\ von Baeyer, \booktitle{QBism:\ The
  Future of Quantum Physics} (Harvard University Press, 2016).
    
\bibitem{Stacey:2019} B.\ C.\ Stacey, ``On QBism and Assumption
  (Q),'' \arxiv{1907.03805} (2019).

\bibitem{Stacey:2016} B.\ C.\ Stacey,
  ``\hrefdoi{10.1098/rsta.2015.0235}{Von Neumann was not a Quantum
  Bayesian},'' \booktitle{Philosophical Transactions of the Royal
  Society A} \textbf{374} (2016), 20150235, \arxiv{1412.2409}.

\bibitem{Garrett:1993} A.\ J.\ M.\ Garrett, ``Making Sense of Quantum
  Mechanics:\ Why You Should Believe in Hidden Variables.''  In
  \booktitle{Maximum Entropy and Bayesian Methods (Paris, France,
    1992)}, edited by A. Mohammed-Djafari and G. Demoment (Kluwer,
  1993).

\bibitem{Ramsey:1926} F.\ P.\ Ramsey, ``Truth and probability.'' In
  \booktitle{The Foundations of Mathematics and other Logical Essays},
  edited by R.\ B.\ Braithwaite (Routledge \& Kegan Paul Ltd., 1931).
  
\bibitem{Jeffrey:1989} R.\ Jeffrey,
  ``\href{https://www.jstor.org/stable/20012238}{Reading
  `Probabilismo'},'' \booktitle{Erkenntnis} \textbf{31} (1989),
  225--37.
  
\bibitem{Stacey:2019e} B.\ C.\ Stacey,
  ``\hrefdoi{10.1387/theoria.20465}{Book Review:\ \booktitle{What Is
    Quantum Information?}},''
  \booktitle{\href{http://www.ehu.eus/ojs/index.php/THEORIA}{Theoria}}
  \textbf{34} (2019), 153--55.

\bibitem{Spekkens:2007} R.\ W.\ Spekkens,
  ``\hrefdoi{10.1103/PhysRevA.75.032110}{Evidence for the epistemic
  view of quantum states: A toy theory},'' \booktitle{Physical Review
  A} \textbf{75,} 3 (2007), 032110, \arxiv{quant-ph/0401052}.

\bibitem{Appleby:2017} M.\ Appleby, C.\ A.\ Fuchs, B.\ C.\ Stacey and
  H.\ Zhu, ``\hrefdoi{10.1140/epjd/e2017-80024-y}{Introducing the
    Qplex:\ A novel arena for quantum theory},'' \booktitle{European
    Physical Journal D} \textbf{71} (2017), 197, \arxiv{1612.03234}.

\bibitem{Conover:2018} E.\ Conover,
  ``\href{https://www.sciencenews.org/article/official-redefining-kilogram-units-measurement}{It's
  official:\ We're redefining the kilogram},'' \booktitle{Science News}
  (16 November 2018).
  
\bibitem{Fuchs:1997} C.\ A.\ Fuchs,
  ``\hrefdoi{10.1103/PhysRevLett.79.1162}{Nonorthogonal quantum states
  maximize classical information capacity},'' \booktitle{Physical
  Review Letters} \textbf{79} (1997), 1162, \arxiv{quant-ph/9703043}.
  
\bibitem{Stacey:2019c} B.\ C.\ Stacey, ``Sporadic SICs and exceptional
  Lie algebras,'' \arxiv{1911.05809} (2019). Originally posted at the
  \booktitle{n-Category Caf\'e}
  (\href{https://golem.ph.utexas.edu/category/2019/02/sporadic_sics_and_exceptional.html}{20
    February},
  \href{https://golem.ph.utexas.edu/category/2019/02/sporadic_sics_and_exceptional_1.html}{28
    February} and
  \href{https://golem.ph.utexas.edu/category/2019/03/sporadic_sics_and_exceptional_2.html}{14
    March 2019}).

\bibitem{DeBrota:2018b} J.\ B.\ DeBrota, C.\ A.\ Fuchs and
  B.\ C.\ Stacey, ``Symmetric Informationally Complete measurements
  identify the irreducible difference between classical and quantum,''
  \arxiv{1805.08721} (2018). Forthcoming in \booktitle{Physical Review
    Research}.

\bibitem{Stacey:2019b} B.\ C.\ Stacey, ``Quantum theory as symmetry
  broken by vitality,'' \arxiv{1907.02432} (2019).

\bibitem{Jaynes:2003} E. T. Jaynes, \booktitle{Probability
  Theory:\ The Logic of Science} (Cambridge University Press,
  2003). Cited in CFS 2002 as unpublished but available at the
  \href{https://bayes.wustl.edu}{Jaynes memorial website}.

\bibitem{Gleason:1957} A.\ M.\ Gleason,
  ``\href{http://www.iumj.indiana.edu/IUMJ/FULLTEXT/1957/6/56050}{Measures
  on the closed subspaces of a Hilbert space},'' \booktitle{Indiana
  University Mathematics Journal} \textbf{6,} 4 (1957), 885--93.

\bibitem{Busch:2003} P.\ Busch,
  ``\hrefdoi{10.1103/PhysRevLett.91.120403}{Quantum states and
  generalized observables:\ A simple proof of Gleason's theorem},''
  \booktitle{Physical Review Letters} \textbf{91} (2003), 120403,
  \arxiv{quant-ph/9909073}.

\bibitem{Caves:2004} C.\ M.\ Caves, C.\ A.\ Fuchs, K.\ K.\ Manne and
  J.\ M.\ Renes,
  ``\hrefdoi{10.1023/B:FOOP.0000019581.00318.a5}{Gleason-type
    derivations of the quantum probability rule for generalized
    measurements},'' \booktitle{Foundations of Physics} \textbf{34}
  (2004), 193--209, \arxiv{quant-ph/0306179}.
  
\bibitem{Hermann:1935} G.\ Hermann, ``\hrefdoi{10.1007/BF01491142}{Die
  Naturphilosophischen Grundlagen der Quantenmechanik},''
  \booktitle{Die Naturwissenschaften} \textbf{42} (1935), 718--21.

\bibitem{Hermann:1935b} G.\ Hermann, ``Die Naturphilosophischen
  Grundlagen der Quantenmechanik,'' \booktitle{Abhandlungen der
    Fries'schen Schule} \textbf{6} (1935), 69--152.
    
\bibitem{Wick:1995} D.\ Wick, \booktitle{The Infamous Boundary:\ Seven
  Decades of Heresy in Quantum Physics} (Copernicus, 1995).

\bibitem{Mermin:2018} N.\ D.\ Mermin and R.\ Schack,
  ``\hrefdoi{10.1007/s10701-018-0197-5}{Homer nodded:\ von Neumann's
  surprising oversight},'' \booktitle{Foundations of Physics}
  \textbf{48} (2018), 1007--20, \arxiv{1805.10311}.

\bibitem{Stacey:2019d} B.\ C.\ Stacey, ``From Gender to Gleason:\ The
  Case of Adam Becker's \booktitle{What Is Real?},''
  \url{https://www.sunclipse.org/?p=2658} (2019).

\bibitem{Granstroem:2006} H.\ Granstr\"om,
  \booktitle{\href{http://kiko.fysik.su.se/en/thesis/helena-master.pdf}{Gleason's
      Theorem}.} Master's thesis, Stockholm University, 2006.
    
\bibitem{Wright:2019a} V.\ J.\ Wright and S.\ Weigert,
  ``\hrefdoi{10.1088/1751-8121/aaf93d}{A Gleason-type theorem for
  qubits based on mixtures of projective measurements},''
  \booktitle{Journal of Physics A} \textbf{52} (2019), 055301,
  \arxiv{1808.08091}.
  
\bibitem{Wright:2019b} V.\ J.\ Wright and S.\ Weigert,
  ``\hrefdoi{10.1007/s10701-019-00275-x}{Gleason-type theorems from
  Cauchy's functional equation},'' \booktitle{Foundations of Physics}
  \textbf{49} (2019), 594--606.

\bibitem{Zurek:2018} W.\ H.\ Zurek, ``Quantum theory of the
  classical:\ Quantum jumps, Born's rule, and objective classical
  reality via Quantum Darwinism,'' \booktitle{Philosophical
    Transactions of the Royal Society A} \textbf{376} (2018),
  20180107, \arxiv{1807.02092}.

\bibitem{Aoyama:2015} T.\ Aoyama, M.\ Hayakawa, T.\ Kinoshita and
  M. Nio, ``\hrefdoi{10.1103/PhysRevD.91.033006}{Tenth-order electron
    anomalous magnetic moment --- contribution of diagrams without
    closed lepton loops},'' \booktitle{Physical Review D} \textbf{91}
  (2015), 033006, \arxiv{1412.8284}.
  
\bibitem{Soler:1995} M.\ P.\ Sol\`er,
  ``\hrefdoi{10.1080/00927879508825218}{Characterization of Hilbert
  spaces by orthomodular spaces},'' \booktitle{Communications in
  Algebra} \textbf{23} (1995), 219--43.

\bibitem{Holland:1995} S.\ S.\ Holland, Jr.,
  ``\hrefdoi{10.1090/s0273-0979-1995-00593-8}{Orthomodularity in
  infinite dimensions; A theorem of M.\ Sol\`er},''
  \booktitle{Bulletin of the American Mathematical Society}
  \textbf{32} (1995), 205--34, \arxiv{math/9504224}.

\bibitem{Hudson:1976} R.\ L.\ Hudson and G.\ R.\ Moody,
  ``\hrefdoi{10.1007/BF00534784}{Locally normal symmetric states and
  an analogue of de Finetti's theorem},'' \booktitle{Zeitschrift f\"ur
  Wahrscheinlichkeitstheorie und Verwandte Gebiete} \textbf{33}
  (1976), 343--51.

\bibitem{Caves:2002b} C.\ M.\ Caves, C.\ A.\ Fuchs and R.\ Schack,
  ``\hrefdoi{10.1063/1.1494475}{Unknown quantum states:\ The quantum
  de Finetti representation},'' \booktitle{Journal of Mathematical
  Physics} \textbf{43} (2002), 4537, \arxiv{quant-ph/0104088}.

\bibitem{Fuchs:2004} C.\ A.\ Fuchs, R.\ Schack and P.\ F.\ Scudo,
  ``\hrefdoi{10.1103/PhysRevA.69.062305}{De Finetti representation
  theorem for quantum-process tomography},'' \booktitle{Physical
  Review A} \textbf{69} (2004), 062305, \arxiv{quant-ph/0307198}.
  
\bibitem{Brandao:2017} F.\ G.\ S.\ L.\ Brand\~ao and A.\ Harrow,
  ``\hrefdoi{10.1007/s00220-017-2880-3}{Quantum de Finetti theorems
  under local measurements with applications},''
  \booktitle{Communications in Mathematical Physics} \textbf{353}
  (2017), 469--506, \arxiv{1210.6367}.
  
\bibitem{Schack:2001} R.\ Schack, T.\ A.\ Brun and C.\ M.\ Caves,
  ``\hrefdoi{10.1103/PhysRevA.64.014305}{Quantum Bayes rule},''
  \booktitle{Physical Review A} \textbf{64} (2001), 014305,
  \arxiv{quant-ph/0008113}.

\bibitem{Stacey:2015} B.\ C.\ Stacey, \booktitle{Multiscale Structure
  in Eco-Evolutionary Ecology.} PhD thesis, Brandeis University,
  2015. \arxiv{1509.02958}.

\bibitem{DeBrota:2019} J.\ B.\ DeBrota and B.\ C.\ Stacey, ``L\"uders
  channels and SIC existence,'' \arxiv{1907.10999} (2019). Forthcoming
  in \booktitle{Physical Review A}.
  
\bibitem{Fuchs:2010} C.\ A.\ Fuchs, ``QBism, the perimeter of Quantum
  Bayesianism,'' \arxiv{1003.5209} (2010).

\bibitem{Appleby:2005} M.\ Appleby,
  ``\hrefdoi{10.1007/s10701-004-2014-6}{Facts, values and quanta},''
  \booktitle{Foundations of Physics} \textbf{35} (2005), 627,
  \arxiv{quant-ph/0402015}.

\bibitem{Fuchs:2009b} C.\ A.\ Fuchs and R.\ Schack,
  ``\hrefdoi{10.1063/1.3109948}{Priors in quantum Bayesian
  inference},'' \booktitle{AIP Conference Proceedings} \textbf{1101}
  (2009), 255--59, \arxiv{0906.1714}.
  
\bibitem{Stacey:2016b} B.\ C.\ Stacey, ``SIC-POVMs and Compatibility
  among Quantum States,'' \booktitle{Mathematics} \textbf{4} (2016),
  36, \arxiv{1404.3774}.

\bibitem{Fuchs:2003} C.\ A.\ Fuchs, \booktitle{Notes on a Paulian
  Idea} (V\"axj\"o University Press, 2003).

\bibitem{Fuchs:1996} C.\ A.\ Fuchs and A.\ Peres,
  ``\hrefdoi{10.1103/PhysRevA.53.2038}{Quantum state disturbance
  vs.\ information gain:\ Uncertainty relations for quantum
  information},'' \booktitle{Physical Review A} \textbf{53} (1996),
  2038--2045, \arxiv{quant-ph/9512023}.

\bibitem{Fuchs:1998} C.\ A.\ Fuchs, ``Information gain vs.\ state
  disturbance in quantum theory,'' \booktitle{Fortschritte der Physik}
  \textbf{46} (1998), 535--565, \arxiv{quant-ph/9611010}.

\bibitem{Fuchs:1998b} C.\ A.\ Fuchs, ``Just \emph{two} nonorthogonal
  quantum states,'' \arxiv{quant-ph/9810032} (1998).

\bibitem{Spekkens:2018} D.\ Schmid and R.\ W.\ Spekkens,
  ``\hrefdoi{10.1103/PhysRevX.8.011015}{Contextual advantage for state
  discrimination},'' \booktitle{Physical Review X} \textbf{8} (2018),
  011015.

\bibitem{Fuchs:2002} C.\ A.\ Fuchs, ``Quantum mechanics as quantum
  information (and only a little more),'' \arxiv{quant-ph/0205039}
  (2002).

\bibitem{DeBrota:2018c} J.\ B.\ DeBrota, C.\ A.\ Fuchs and
  B.\ C.\ Stacey, ``Analysis and synthesis of Minimal Informationally
  Complete quantum measurements,'' \arxiv{1812.08762} (2018).

\bibitem{Riehl:2016} E.\ Riehl,
  \href{http://www.math.jhu.edu/~eriehl/context.pdf}{\booktitle{Category
      Theory in Context}} (Dover, 2016).

\bibitem{Barnum:2010} H.\ Barnum, S.\ Beigi, S.\ Boixo,
  M.\ B.\ Elliott and S.\ Wehner,
  ``\hrefdoi{10.1103/PhysRevLett.104.140401}{Local quantum measurement
    and no-signaling imply quantum correlations},''
  \booktitle{Physical Review Letters} \textbf{104} (2010), 140401,
  \arxiv{0910.3952}.
  
\bibitem{DeLaTorre:2012} G.\ de la Torre, L.\ Masanes, A.\ J.\ Short
  and M.\ P.\ M\"uller,
  ``\hrefdoi{10.1103/PhysRevLett.109.090403}{Deriving quantum theory
    from its local structure and reversibility},'' \booktitle{Physical
    Review Letters} \textbf{109} (2012), 90403, \arxiv{1110.5482}.

\bibitem{Fuchs:2012} C.\ A.\ Fuchs and R.\ Schack, ``Bayesian
  conditioning, the reflection principle, and quantum decoherence,''
  \booktitle{Probability in Physics} (2012), 233--47,
  \arxiv{1103.5950}.

\bibitem{Vaidman:2019} L.\ Vaidman, ``Derivations of the Born Rule,''
  \url{http://philsci-archive.pitt.edu/15943/} (2019).

\bibitem{Laudisa:2019} F.\ Laudisa and C.\ Rovelli,
  ``\href{https://plato.stanford.edu/entries/qm-relational/}{Relational
  Quantum Mechanics},'' \booktitle{Stanford Encyclopedia of
  Philosophy} (2019).

\bibitem{Jaeger:2019} G.\ Jaeger,
  ``\hrefdoi{10.1002/andp.201800097}{Information and the
  reconstruction of quantum physics},'' \booktitle{Annalen der Physik}
  \textbf{531} (2019), 1800097.

\bibitem{Fuchs:2000} C.\ A.\ Fuchs and A.\ Peres,
  ``\hrefdoi{10.1063/1.883004}{Quantum theory needs no
  `interpretation'},'' \booktitle{Physics Today} \textbf{53} (March
  2000), 70--71.
    
\bibitem{Mermin:2018b} N.\ D.\ Mermin,
  ``\hrefdoi{10.1088/1361-6633/aae2c6}{Making better sense of quantum
  mechanics},'' \booktitle{Reports on Progress in Physics} \textbf{82}
  (2018), 012002, \arxiv{1809.01639}.

\bibitem{Peierls:1991} R.\ Peierls,
  ``\hrefdoi{10.1088/2058-7058/4/1/19}{In defence of `measurement'},''
  \booktitle{Physics World} \textbf{4} (January 1991), 19--20.

\bibitem{ONeil:2016} C.\ O'Neil, \booktitle{Weapons of Math
  Destruction} (Crown Books, 2016).
  
\end{thebibliography}
\end{document}